\documentclass[conference]{IEEEtran}
\IEEEoverridecommandlockouts
\usepackage{cite}
\usepackage{amsmath,amssymb,amsfonts}
\usepackage{algorithmic}
\usepackage{graphicx}
\usepackage{textcomp}
\usepackage{array}

\usepackage{xcolor}
\def\BibTeX{{\rm B\kern-.05em{\sc i\kern-.025em b}\kern-.08em
    T\kern-.1667em\lower.7ex\hbox{E}\kern-.125emX}}
\begin{document}

\title{Ontology Based Information Integration: A Survey
}

\author{\IEEEauthorblockN{1\textsuperscript{st} Maryam Alizadeh}
\IEEEauthorblockA{\textit{Pouyandegane-danesh }\\\textit{higher education institute} \\
Chalus, Iran\\
marya.alizade@pd.ac.ir\\
}
\and
\IEEEauthorblockN{2\textsuperscript{nd} Maliheh Heydarpour Shahrezaei}
\IEEEauthorblockA{\textit{Pouyandegane-danesh }\\\textit{higher education institute} \\
Chalus, Iran \\
maliheh.heydarpour@pd.ac.ir\\
}
\and
\IEEEauthorblockN{3\textsuperscript{rd} Farajollah Tahernezhad-Javazm }
\IEEEauthorblockA{\textit{Department of Mechatronics} \\
\textit{University of Tabriz}\\
Tabriz, Iran\\
Tahernejad92@ms.tabrizu.ac.ir\\
}
}

\maketitle

\begin{abstract}
An ontology makes a special vocabulary which describes the domain of
interest and the meaning of the term on that vocabulary. Based on the precision of
the specification, the concept of the ontology contains several data and conceptual
models. The notion of ontology has emerged into wide ranges of applications
including database integration, peer-to-peer systems, e-commerce, semantic web,
etc. It can be considered as a practical tool for conceptualizing things which are
expressed in computer format. This paper is devoted to ontology matching as a
mean for information integration. Several matching solutions have been presented
from various areas such as databases, information systems and artificial intelligence.
All of them take advantages of different attributes of ontology like, structures, data
instances, semantics and labels and its other valuable properties. The solutions have
some common techniques and cope with similar problems, but use different methods
for combining and exploiting their results. Information integration is among the first
classes of applications at which matching was considered as a probable solution.
Information integration contains many fields including, data integration, schema
integration, catalogue integration and semantic integration. We cover these notions
in term of ontology in our proposed paper.
\end{abstract}


\section{Intoroduction}
\label{sec:1}

To integrate information across different resources, it is important to have a
formalization of the mental concepts that people have about different entities. In term
of information integration, ontology is used to identify correspondence between
entities of the local information sources which are semantically related.  Information integration is categorized as the first classes of topics which
considers ontology matching as plausible solution.

In general, an information integration system includes multiple steps:
\begin{itemize}
\item Translate the query in terms of the common ontology
\item Identify the correspondence among entities of the local sources of
information and the common ontology which are semantically related to each
other
\item Interpret the data instances of local information sources into a knowledge
representation of the information integration system
\item Adopt the results of different information sources and eliminating the
redundancies before representing the last answer.
\end{itemize}

Information integration contains many fields including, data integration \cite{2}, schema
integration \cite{3}, catalogue integration \cite{4} and semantic integration \cite{5}. In the
following we provide the definitions of the prementioned concepts:
\begin{itemize}
\item Schema Integration: In this integration scenario, if multiple enterprises want
to merge, they need to identify the correspondences between different
entities of schema before doing the merge processes. This is called matching
process and is required if different databases want to be combined.
\item Catalogue Integration: In B2B or Business to Business applications, the
partners store their information in the form of electronic catalogues. If a
merchant wants to be member of a market, the correspondences between the
entities of its catalogue and market catalogue should be first determined. The process of finding correspondences between catalogues is called catalogue
matching.
\item  Data Integration: Data integration is a method at which information from
multiple data sources are integrated without loading data into a central
storage. This helps all the local sources to have access to up-to-date data and
information. The point is that the central storage should be updated by data
integration system.
\item Semantic Integration: Semantic integration is the process of incorporation
detailed semantics about data which provide more consistency when using
and understanding it. The biggest benefit of that is the reduction of human
involvement for data integration and data interesting. Semantic integration
does not just consider organizing the data but it further provides more about
their concepts.

\end{itemize}

In this paper we present a survey of existing solutions of ontology-based information integration. We also discuss some points and concepts in the ontology based integration systems. The  paper is organized as follows: In section \ref{sec:2}, we discuss the role of ontologies. In section \ref{sec:M}, we describe the notion of use of mapping in ontology systems. Section \ref{sec:3}, introduces the ontology evaluation.  We provide the survey of existing works in section \ref{sec:4}. The summary of the state-of-the-art is purveyed in  section \ref{sec : dis}. Finally, we conclude the paper at section \ref{sec:con}.

\section{The Role of Ontologies}
\label{sec:2}

Ontologies can be used in different area of studies and can be used in different area of the computer science  like security, intelligent systems, etc. In security, they can be used to secure the networks and to substitute with other methods of securing the networks and computer systems \cite{6, 7, 8, 9, 20}. They can even be utilized for optimizing the life of the networks to further improve the methods discussed in \cite{10}.  As previously discussed, ontologies also can be used for integration task to describe and
define the semantics for information sources and to explain the semantics of those
resources and make its context explicit. Ontologies can also be extended for other
applications in other projects which we describe as follow:

\begin{itemize}
\item Content Explication: Ontologies are introduced as explicit specification of a
conceptualization. They can be used for identification of semantically related
information concepts.
\item Single ontology approaches: Use one global ontology which provide shared
vocabulary for specification of the semantics. Single ontology methods can be
used for integration problem at which we have the same view on the domain.
\item Multiple ontology approaches: In this method, each information source has its
special ontology. In this method, there is no need to commitment to a global
ontology. Each source ontology is developed without respect to other sources
and their ontologies.
\item Hybrid approaches: Similar to hybrid method, the semantic of each source is
defined with its own ontology. But, local ontologies are built from global
shared vocabulary in order to be comparable to each other.
\end{itemize}

In the following we describe the use of mappings in the ontology context.

\section{Use of Mappings}
\label{sec:M}
Integrating heterogeneous information from different data resources is a big
challenge. Ontologies need to be related to their environment and this plays an
important role in information integration. Mapping refers to the connection of an ontology to other parts of the application. In case of mapping, several concepts should
be considered which are as follow:

\begin{itemize}
\item Connection to Information Sources: \\
First and for most, ontologies should be related to database scheme. Several
methods are used for this:\\
\textbf{ Structure resemblance
}In this method, there is a simple one to one copy of the structure of the
database and it is encoded to a language for making automated reasoning
possible. The integration is done in copy of the model.
\textbf{Structure enrichment}:\\ A common approach for relating ontologies to
information resources. It uses the structure resemblance and a more
definition over concepts and terms.

\item Inter Ontology Mapping:\\
Many methods use more than one ontology for describing the information. Then
there should be a method for mapping different ontologies.\\
\textbf{Defined Mappings:} In this method, translation between ontologies are done
by special agent which has the role of a mediator and tries to translate
between different ontologies and different languages. In this method,
different methods of mappings including one-to-one mappings between the
values and mapping between compound expressions is considered. This
method has an outstanding flexibility but no observation over semantics since
user can freely define arbitrary mappings even if they are meaningless.

\textbf{Lexical relations:} In this method, mapping between concepts of different
ontologies is provides. In this method, relationships can be defined as
synonyms, hypernyms, overlaps, coverings and disjoints. They are easy to be
constructed but they do not have formal semantics.
\textbf{Top-Level Grounding:} If one wants to not lose the semantics, there should
be formal representing language when mapping between different ontologies
occur. In order to do this, a single top-level ontology should be defined. This
method resolves the conflicts and ambiguities. This method can establish
connections between concepts of the all ontologies but can cause problems
for matching if not establish a direct correspondence.

\textbf{Semantic Correspondence: }This approach tries to amend the problem of
ambiguity. Ambiguity is produced when indirect mapping of concepts from a
top-level grounding tries to identify semantic correspondences of different
concepts of ontologies. For avoiding this, approaches should more rely on a
specific common vocabulary in order to be able to define concepts across
different ontologies. Semantic correspondences can ameliorate this problem.

\end{itemize}

In the following subsection, we describe the methods of evaluation in the ontology.
\section{Ontology Evaluation}
\label{sec:3}

When developing ontologies for integration systems, one should consider that
ontologies evolve regularly. Therefore, integration systems must be flexible when
sources departure and arrival occur. They also should be able to handle any changes
in information sources. Typical solution would be to regenerate the mappings. The
approach, to recreate mappings whenever ontology changes is not unproblematic
and each time the previously captured information should be utilized. The big
challenge of data integration with evolving ontologies is the issue of mapping
adaptation. So one may do mapping adaptation each time to deal with ontology
evolution. However, in both cases of mapping regeneration and mapping adaptation,
semantic relation of ontologies might be lost. Therefore, new methods should
consider this issue into mind. There are few works which tackle it but not all of the
approaches cover this problem.
In our survey project, we will provide a detailed discussion of these concepts and will
talk about common methods in ontology based integration field and the technical
solutions that these papers have proposed. As part of our motivation, we think that
this survey project can be used as a best way to get an idea about the recent
technologies, methodologies, algorithms and tools in the field of ontology matching
for information integration and one can get a clear overview of the current state of
this field. The project can be served as a great starting point for researchers and
academics to think about venturing into this field. They can also easily take updated
information about semantic integration and ontology based information integration
and to be familiar with challenges, gaps, unanswered questions and future works of
this area of research. At last, the major difference between each work would be
purveyed.
 
\section{Related Work}
\label{sec:4}
In this survey, we consider overviewing the state of art in ontology-based information integrity like what is proposed in \cite{11, 12, 13}. These days, there are vast demands for developing techniques and methods which
can process complicated data and simplify efficient data interoperability. Among
different methods, ontology has substantiated that can handle data heterogeneity. In
most of the ontology based integration systems, there is a global ontology which is
integrated view of the data sources. Since developing and constructing ontologies is
not feasible in all the domains, this paper has presented semi-automated method for
developing ontology. Their approach is based on Formal Concept analysis (FCA)
which can deal with ontology development by abstracting conceptual structures from
attribute-based object and can automate the ontology development activities. In the
proposed methods, authors make a classical FCA theory to develop ontology for
integrating datasets which include implicit and ambiguous data. Implicit data causes
a not well-formed ontology which cannot support critical concepts and semantics of
them. Ambiguous data eventuates in inconsistent between different datasets. In this
paper, the author has considered implicitly and ambiguity as key factors for
developing ontologies. In their works, they have devised some rules for restoring
implicit information. They have provided several examples on how implicit data
cannot be retrieved. 
To resolve disembogues data, they invent a list of basic operations, for finding simple
match and then further processed for dealing with more complicated matches.
As previously mentioned, in the paper it has been mentioned that implicit information
caused ill-formed ontology. In the paper, some rules have been extracted and made. 
For finding the rules, they iterate over the attributes. Attributes have information if
there is missing value for object. Experts should also involve in finding and making
the rules. Objects rules are derived by iterating over the objects and searching for the
attributes which do not have an exclusive value. If attributed have implicit
information which cannot be recovered with attribute rules, an object rule will
recover this information by getting help from experts. Then the paper explains how
the rules are developed. Once the rules have been identified, new objects will be
generated by applying different combination of the rules. This help objects which
have different combination of attributes to be extracted. They then provide a table
consists many valued contexts after restoring implicit information with rules. the
many values context will be then fed into the Conceptual Scaling component which
can generate a one valued context table.
In their infrastructure, there is also Context Composition Component which is
responsible for taking two contexts as input and make an integrated GSH as an output.
This component includes two main components: Context Integration and Hierarchy
Generation. This component should deal with ambiguous information when it is
integrating the context. For attribute disambiguation, the paper use a pre-defined
data dictionary to disambiguate the attributes. Using dictionary, they will be able to
find semantic relationships between attributes. Then they identify semantic
relationship between attributes and for each attribute a semantic mapping operation
with all the attributes would be done. Based on the mapping which has been done for
a special attribute, new attributes and relationships will be added. Each attribute can
find mapping of several types (one to one or one to many). 
For each of the attributes, four diverse types of mappings are derived:
\begin{itemize}
\item Attribute ($A_i$) finds equivalent attribute $A_j$. In this case, they will be unified.
\item Attribute ($A_i$) finds a match attribute $A_j$ which is more generic than it. In this
case the resulting context in a special set $K_1$ is expanded by $A_i$ and
relationships of $A_i$ with objects in set $K_2$.
\item Attribute $A_i$
finds a match to attribute $A_j$ which is more specific to it. For this
case, the context of set $K_1$ is extended with $A_i$ and existing relationships
between $A_i$ and objects of set $K_2$.
\item Attribute $A_i$ does not find any match in the set $K_1$. In this case, the context of
$K_1$ is just extended by $A_i$ and existing relationships between it and other
objects which are basically originated from set $K_2$.

\end{itemize}

If there are multiple matches of distinct types, the primitive operations which were
discussed can simply be composed to deal with these complex cases. This procedure
has been nominated as "Composite Operations" and is explained in the paper within
some examples. In this paper, ontology derivation part has different components
which takes the GSH and results in an ontological structure. The GSH is responsible
for deriving different information including ontological concepts, relationships
between concepts and attributes of the concepts. The components of the ontology
derivation are as follows:

\begin{itemize}
\item Mapping identification: Once the mapping of different objects have been
identified, it should be validated by experts. If the identified mapping is not
correct, features should be identified and derived to explain the differences
between one concept from the other one. This eventuates in identification of
new attributes or new relationships. In this paper, this has been called as
Mapping Identification.
\item Concept/relationship/attribute identification: In this paper, it has been
demonstrated that all of the intermediate and abstract concepts are
summarized in integration step and only objects concepts and attribute
concepts remain in the resulting hierarchy. All the objects keep in the
ontology and the existence of attributes in the ontology depends on expert
decision. After it is determined that which concepts need to be kept, the rules
will be derived for finding the relationships of all attributes and concepts. 
\end{itemize}
For the evaluation process, this paper has used the data sets of the UK water
companies and the evaluation has been done in two distinct levels including lexical
and taxonomic level. In the lexical level, the evaluation examines if the lexical terms
of source ontology cover the lexical terms in the destination ontology. In the
Taxonomic level, the ontology measures if conceptual hierarchy of the source
ontology resembles the target ontology precisely. The experimental results of this
paper, shows that the techniques on this paper accurately can help for organizing and
merging data of different data sources. The results also precisely demonstrate that
the techniques can support the development of the ontological which more efficiently
can respect the underlying knowledge structure of the domain.

Kondylakis, et. al \cite{12}, analyses the necessity of having an ontology evolution to the integration
system. The biggest challenge that ontologies face is that, ontologies are frequently
changed. When the change occurs, the mappings are not further valid and need to be
updated. The traditional methods used mapping adaption methods for tackling this
problem. But these methods cannot guarantee that the semantics of resulting
mappings remain desirable. To ameliorate this condition, the paper has presented
"Exelixis" as a platform for query answering over evolving ontology without any need
for mapping redefinition. "Exelixis“ system is based on dependent queries. In this
system, data integration is done based on ontology evolving without any mapping
redefinition. To do this, this system rewrites the queries among ontology versions. At
the first step, changes are automatically detected and interpreted as global-as-view
(GAV) mappings.

\begin{figure}
\centering
\includegraphics[width=80mm]{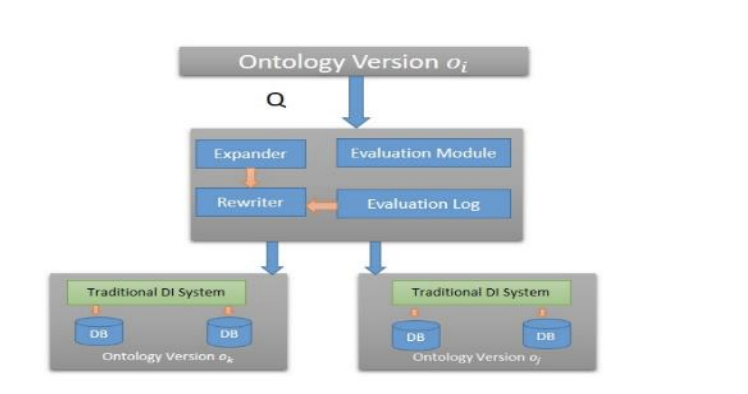}
\caption{Query rewriting in Exelixis.}
\label{Fig2b}
\end{figure}

In the second step, an expansion over the query is applied to meet the constraints
from the ontology. If equivalent rewritings are not available, it tries to guide query
redefinition or over the best over-approximations. The  Fig. \ref{Fig2b} demonstrates the
infrastructure of the proposed work. Generally, their architecture contains three
main components including, "Expander", "Valid Rewriter" and “Change Path
Generator”. "Expander" is responsible for identifying subClass and subProperty of
ontology and tries to rewrite the query based on them. The Valid Rewriter utilizes the
GAV mappings in order to rewrite queries among ontologies.The last component
allows user to be able to solve evolution of the ontology only for a specific part of that.
This could be simply done by "Change Path Generator" which calculate the change
paths for a specific class.
Although, Exelixis is a good framework for solving the problem of ontology evolving,
it is hard to extend this framework to other applications efficiently. Moreover, this
article has not introduced any evaluation system to analyze the efficiency of the
presented system. They have just a demonstration with CIDOC-CRM ontology.

Next to this, Gagnon  in \cite{13} presented a new ontology based on information integration system
with ontology mapping. This work has been proposed as a methodology for
integrating of heterogeneous data sources. Since data and information come from
variety of resources they must be merged, corrected and aggregated together. The
most important purpose of this paper is to develop a system for integrating multiple
data sources efficiently and it focuses on improving automation of integration of data
sources by their ontology. 

\begin{figure}
\centering
\includegraphics[width=80mm]{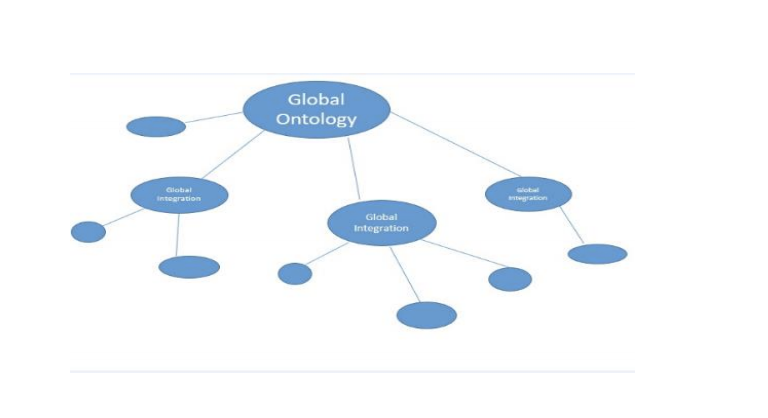}
\caption{ Architecture for semantic integration of data sources.}
\label{Fig3b}
\end{figure}

The provided ontology-based integration system constructs a global ontology from
local ontologies of data sources. Then the data integration system exploits the global
ontology and its integration to the local ontologies of each data source. Considering
the preceding issues, the paper has proposed its main infrastructure. In the
architecture, there is a virtual database which moves a copy of data from several data
sources of a special database. Moreover, a mediator maps the requests and its
correspondence answers between the global and the local ontologies. The
contribution of their method is that there is no need for commitment to a global
ontology. In fact, the local ontologies represent vocabularies which are used in the
same domain to make the synonym relations. As a result, this method is less
consuming than global schema matching method since there needs a few number of
rules and relationships to be defined. They provide their architecture in  Fig.\ref{Fig3b}.
Although this work has proposed a comprehensive model for integrating data
resources, it still does not have an evaluation system for measuring its efficiency and
accuracy. Also, while this technique helps to get a big sketch from global ontology, it
does not use the strengths of schema mapping.

Padilha, et. al \cite{14} presented an ontology alignment for finding data which are related to each other in
multiples ontologies. This concept excessively emerges in semantic heterogeneity.
Generally, semantic integration is to ensure that only data of same real world entity
are emerged and combined. Ontology alignment method can generally be applied for
integrating data in the semantic level. Considering the preceding fact, this paper has
used foundational ontology to improve ontology alignment. In the paper, they also
introduce OntoUML which is a conceptual modeling language for complying with
ontological distinctions and axiomatic theories which is designed by the University Foundational Ontology(UFO). UFO, is a foundational ontology which has been
extended and developed based on the number of theories which are from Formal
Ontology, Philosophical Logics, Philosophy of Language and Linguistic and Cognitive
Psychology. The OntoUML also can be used to make the distinction between objects
and processes, types and their roles, etc. In the paper, it has also been illustrated that
using OntoUML can efficiently improve ontology alignment process for data
integration in the semantic level. Throughout the work, four types of ontology have
also been introduced including Top-level ontologies, domain ontologies, task
ontologies and application ontologies. Top-level ontologies explain general concepts,
Domain ontologies describe the vocabulary which is related to a generic domain, Task
ontology is used for showing a generic task or activity and application ontologies are
for explaining concepts depending on a particular task or domain. They have also
explained about foundational ontologies which present formal semantics for highlevel categories. This kind of ontology is served as a conceptual basic for domain
ontologies. In this paper, it has been mentioned that, the techniques which are
described in the ontology alignment, categorized into element-level (based on the
granularity of the analysis) and terminological, structural, extensional or semantic
(based on the type of the input.). The paper also mentioned about two design patterns
of the ontological foundations for OntoUML which consists The Role Design Pattern
and The Phase Design Pattern.

   \begin{table*}[t]
\centering
\caption{Comparison of the proposed methods.}

\begin{tabular}{ |p{1.5cm}||p{1cm}|p{3cm}|p{3cm}| p{1cm}|p{2cm}|p{2cm}| }
 \hline
&\centering Year&\centering Ontology&\centering Mapping&\centering Evolution&\centering Integration&Evaluation\\
\hline
\centering FCA Method   & \centering 2016&\centering Global (single) Ontology& Inter Ontology
\centering Mapping
(Semantic
Correspondence)/All    &\centering SEMI& \centering Data/Semantic&  Lexical

and
Taxonomic
Level\\
\hline
 \centering Mansukhlal
et.al
Method&  \centering 2016  &\centering Global (single)
Ontology   &\centering Connection to info
Resources(Structure
Resemblance) & \centering NO & \centering Semantic&No\\
\hline
 \centering Padilha
et.al
Method &\centering 2012 & \centering Multiple
Ontologies&  \centering Inter Ontology
Mapping (Lexical
Relations)& \centering NO&\centering Semantic&No\\
\hline
 \centering Exelixis    &\centering 2011 &\centering Multiple
Ontologies& \centering Connection to
Information Sources
(Structure
resemblance)
GAV mapping&\centering YES&\centering  All& CIDOCCRM

Ontology\\
 \hline
\centering  Gagnon
Method& \centering 2007  & \centering Global (single)
Ontology&\centering Inter Ontology
Mapping
(Semantic
correspondence)&\centering YES&\centering Semantic&NO\\
 \hline
\end{tabular}
\label{table1}
\end{table*}

\begin{itemize}
\item Role Design Pattern:

A role can process a meta-property which is named Relational Dependence.
As a result, a OntoUML should always have as super type a kind and should be
connected to a community which represent this relational dependence
condition. In this case, there is a problem which is called the problem of role
with multiple disjoint allowed types
\item Phase Design Pattern:

A phase indicates the phased sortals phase. In this case, the parts are disjoint
and are also complete.
\end{itemize}

\begin{figure}
\centering
\includegraphics[width=80mm]{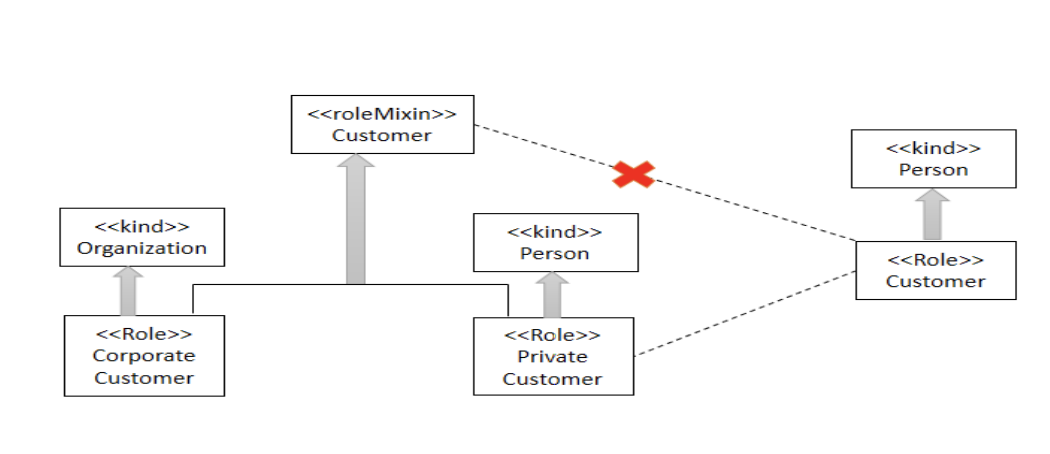}
\caption{ Role Design Pattern.}
\label{Fig4b}
\end{figure}

\begin{figure}
\centering
\includegraphics[width=80mm]{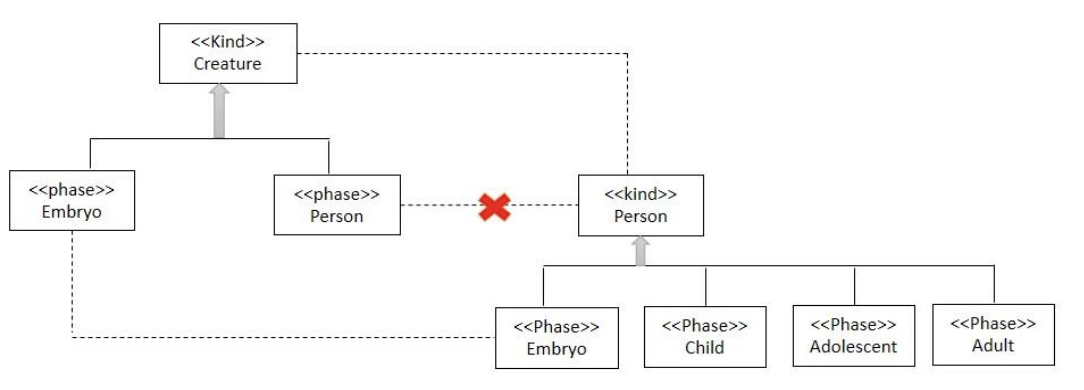}
\caption{ Phase Design Pattern.}
\label{Fig5b}
\end{figure}

\begin{figure}
\centering
\includegraphics[width=80mm]{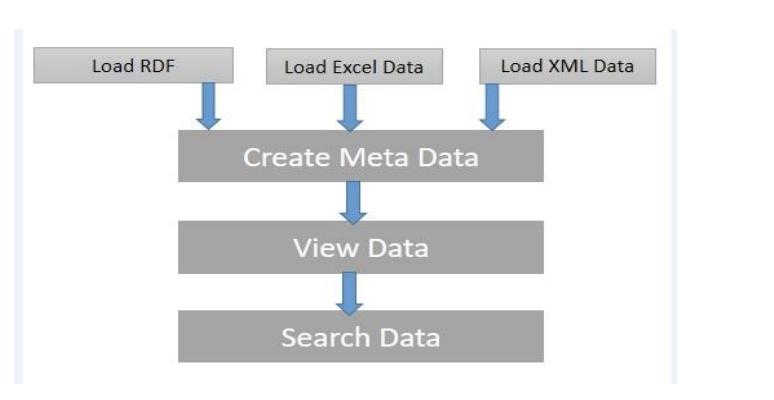}
\caption{  Semantic Integration Infrastructure.}
\label{Fig6b}
\end{figure}

The Role Design Pattern in Fig. \ref{Fig4b} persuades the modeler to provide explicit design
feature which are implicit in a UML model.
As a result, to be sure about the semantic correctness of two different ontologies, one
has to check the principal of the identity and its relational dependence condition
which is explicit in the model. The Phase Design Pattern in Fig.\ref{Fig5b} helps the modeler to
make all of the parts of the phase partition to be explicit. Phases are intrinsically
independent. So, they have considered that if two different kinds are aligned and both
have phases which refer to the same properties intrinsically, then the phases should
be aligned together. In the other case, if the phases mention to the same intrinsic
property but their alignment is different, then the alignment between them is
incorrect.

Overall, in this paper the authors demonstrated that how the use of OntoUML can
ameliorate the alignment process by solving semantically ambiguity by explicit meta
data. However, they did not provide a system for evaluating this approach.

 Manshukhal, et.al \cite{15},  provided a survey for integrating databases by using ontology. It also
has presented a new method in term of integration of databases which can find
dependencies between ontologies. The paper first mentions the meaning of ontology
and its necessity for data integration. In the next parts, it talks about semantic web
and structure of OLAP systems. Authors also have proposed an architecture for
semantic integration. In their system, there are four modules including two modules
for Loading Data and for creating Meta Data and two modules for Viewing and
Searching the Data. The module for loading data allows loading different data formats
including XML, RDF and Excel file. In the Meta Data module a common meta data is
created from combining and merging the different data sets. View Data module
indicates the Meta Data information and Search Data searches information for Meta
data. Their overall structure for semantic integration is provided at Fig.\ref{Fig6b}. Although, this work has claimed about providing a survey for information integration,
there are not outstanding descriptions and explanations about recent works.
Moreover, there is not any evaluations system for analyzing their approach.
In this section, we review the state of art in ontology based information integration.
In section.4 we provide comparison between all of these methods.

\section{Discussion}
\label{sec : dis}    

Based on definitions in section \ref{sec:1}, we classify the papers which is shown in Table.~\ref{table1}.
This table clearly compares the idea and methodology of the five papers reviewed in
this survey. As it is indicated, among all of the proposed methods, only the "Exelixis"
method provided ontology evolution approach in their work. As we previously
mentioned, ontology evolution is important, since ontologies are live artifacts. "FCA"
method also tries to have a methodology for ontology evolution, but it does not
provide this in a perfect manner. Furthermore, just these two approaches evaluate
their proposed methods and others do not have a special system for evaluating their 
systems. In case of ontology model, FCA, Mansukhlal and Gagnon methods use Global
(Single) ontology, while Exelixis and Padilha utilize multiple ontology approach.

As it is illustrated in the Table.~\ref{table1}, FCA, Padilha and Gagnon design consider inter
ontology mapping for mapping procedure and Exelixis and Mansukhlal bring
connection to info resources into their infrastructure. Finally, in case of integration,
we can see that almost all of the state of arts provide semantic integration. However,
Exelixis provide an architecture to deal with all types of information integration.

\section{CONCLUSIONS AND FUTURE WORK}
\label{sec:con}

Information Integration is a big challenge in recent days. Data are from different
resources and need to be collected, merged and fused to be applicable. Overall,
information come from different resources and need to be collected, merged,
corrected and aggregated. Moreover, for making cooperation between divergent 
applications, data sources of any kind should be linked and aggregated. Advanced
information integration systems should also support data fusion and text mining and
should have a potential for handling continuous change and evolution. As a result,
ontology emerged as a suitable tool for this purpose. In this article, we make a brief
overview of most recent works in terms of ontology based information integration
and we provide a comparison between these methods. As future research, we think
about two research fields: First, providing methods and approaches for dealing with
incomplete information for data integration. Second, developing a system for
validating models with capability of making ontology automatically on different
datasets. These two concepts have not been discussed on recent works for ontology
integration systems and are open to debate.

\end{document}